\documentclass[prb,twocolumn,showpacs]{revtex4}
\usepackage{epsfig,amsmath}

\begin{document}

\title{Study of interacting electrons in graphene under the renormalized-ring-diagram approximation} 

\author{Xin-Zhong Yan$^{1,2}$ and C. S. Ting$^1$}
\affiliation{$^{1}$Texas Center for Superconductivity, University of 
Houston, Houston, Texas 77204, USA\\
$^{2}$Institute of Physics, Chinese Academy of Sciences, P.O. Box 603, 
Beijing 100080, China}
 
\date{\today}
 
\begin{abstract}
Using the tight-binding model with long-range Coulomb interactions between electrons, we study some of the electronic properties of graphene. The Coulomb interactions are treated with the renormalized-ring-diagram approximation. By self-consistently solving the integral equations for the Green function, we calculate the spectral density. The obtained result is in agreement with experimental observation. In addition, we also compute the density of states, the distribution functions, and the ground-state energy. Within the present approximation, we find that the imaginary part of the self-energy fixed at the Fermi momentum varies as quadratic in energy close to the chemical potential,  regardless the system is doped or not. This result appears to indicate that the electrons in graphene always behave like a moderately correlated Fermi liquid.
\end{abstract}


\maketitle
\section {Introduction}

The graphene is a single-layer honeycomb lattice of carbon atoms coated on the surface of some materials.\cite{Novoselov,Zhang} The Dirac cone structure in the energy spectrum is responsible for some of the unusual properties of the system.\cite{Semenoff,Fradkin,Haldane} Study of the behaviors of electrons in graphene is one of the currently focused areas in the condensed-matter physics. In the theoretical investigations, most of the calculations are based on the continuous model with simplified Dirac cone dispersion,\cite{Sarma,Barlas} and the Coulomb interactions between electrons are treated in the random-phase approximation (RPA). Since the interactions are correctly taken into account in the long wavelength limit, this approach should reasonably describe the low energy behaviors of the electrons within the validity of RPA. In such a model, however, the lattice-structure effect and the short-range part of the Coulomb interaction have been completely neglected. Thus, it is desirable to explore this problem by a more realistic approach including the effect of the graphene lattice and a self-consistent scheme beyond the RPA. 

In this paper, we use the tight-binding model defined on the two-dimensional honeycomb lattice to formulate the Green function theory of electrons in graphene. In the self-energy of electron Green function, the Coulomb interactions are taken into account with the renormalized-ring-diagram approximation (RRDA). This approximation is well known to satisfy the microscopic conservation laws.\cite{Baym} Our recent investigation of the two-dimensional electron system\cite{Yan1} shows that the RRDA can accurately reproduce the result of the fixed-node-diffusion Monte Carlo simulation for the ground-state energy.\cite{Attac} It is therefore expected that the RRDA could give more reliable description for the behaviors of electrons in graphene.

\section{Lattice Structure and Fourier Transform}

For readers' convenience, we here briefly review the structure of a honeycomb lattice and its reciprocal lattice.\cite{Wallace,Slonczewski} For the sake of numerical computation, we will also present the mapping between coordinates defined on the basic vectors of the honeycomb lattice and the orthogonal coordinates.   

\begin{figure} 
\centerline{\epsfig{file=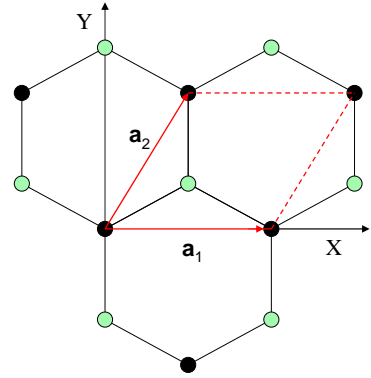,width=6.2 cm}}
\caption{(Color online) The structure of a honeycomb lattice. The basic vectors of the lattice are {$\bf a_1$} and {$\bf a_2$}. There are two sites in each unit cell enclosed by the red lines (solid and dotted): black and green.}\label{fig1}
\end{figure}

\begin{figure} 
\centerline{\epsfig{file=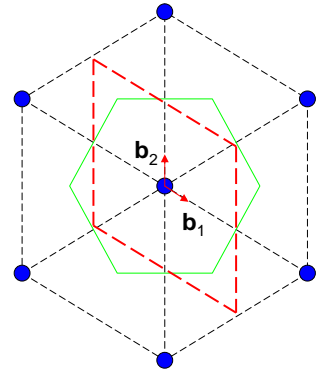,width=6.5 cm}}
\caption{(Color online) Reciprocal lattice of honeycomb structure. The basic vectors of the lattice are {$2\pi\bf b_1$} and {$2\pi\bf b_2$}. The first Brillouin zone is the hexagon with the green boundary. The diamond enclosed by the red dashed lines is the equivalent first Brillouin zone.}\label{fig2}
\end{figure}

The graphene lattice is of the honeycomb structure as shown in Fig. \ref{fig1}. A set of basic displacement vectors of the lattice is 
\begin{eqnarray}
{\bf a_1} &=& (1,0)a \\
{\bf a_2} &=& (\frac{1}{2},\frac{\sqrt{3}}{2})a  
\end{eqnarray}
where $a$ is the lattice constant. We will chose $a$ as the unit of length, and thereby set $a = 1$ hereafter. The area of the unit cell is 
\begin{equation}
S = \frac{\sqrt{3}}{2}
\end{equation}
in unit of $a^2 = 1$. The whole lattice can be viewed as a quadrilateral lattice consisting of the unit diamond cells. There are two sites in each unit cell: black and green. With {$\bf a_1$} and {$\bf a_2$}, we then define the unit vectors of the reciprocal lattice shown in Fig. \ref{fig2}. They are given by
\begin{eqnarray}
{\bf b_1} &=& {\bf a_2 \times z}/S = (1,-\frac{1}{\sqrt{3}}) \\
{\bf b_2} &=& {\bf z \times a_1}/S = (0, \frac{2}{\sqrt{3}})  
\end{eqnarray}
where {\bf z} is the unit vector in the direction of {$\bf a_1 \times a_2$}. The basic displacement vectors of the reciprocal lattice are $2\pi\bf b_1$ and $2\pi\bf b_2$. The first Brillouin zone (BZ) is the hexagon. The diamond enclosed by the red dashed lines in Fig. \ref{fig2} is an equivalent first BZ that is a convenient choice for numerical calculation. 

For the use of numerical calculation, we here write down the transform between the coordinates on the basis of \{{$\bf a_1, a_2$}\} and the orthogonal axes in real space. Consider a vector $\vec r = (x,y)$ in the representation of \{{$\bf a_1, a_2$}\}. We denote this vector in the orthogonal coordinate system as $\vec R = (X,Y)$. Then the correspondence between these two sets of coordinates is given by
\begin{eqnarray}
X &=& x+\frac{y}{2} \\
Y &=& \frac{\sqrt{3}}{2}y.  
\end{eqnarray}
The matrix $\Hat T$ of the transform $\vec R = \Hat T \vec r$ is therefore given by
\begin{equation}
\Hat T = 
\begin{pmatrix} 
1& \frac{1}{2} \\\\
0& \frac{\sqrt{3}}{2}
\end{pmatrix}
\end{equation}
where the first and second columns are the coordinates of vectors $\bf a_1$ and $\bf a_2$ respectively. Analogously, in the momentum space, we obtain the transform between the coordinates of a vector $\vec Q$ defined in the orthogonal system and its projections $\vec q$ on the basis \{$\bf b_1, b_2$\}. Denoting the transform as $\vec Q = \Hat M \vec q$, we have 
\begin{equation}
\Hat M = \begin{pmatrix} 1& 0 \\\\
-\frac{1}{\sqrt{3}}& \frac{2}{\sqrt{3}} 
\end{pmatrix},
\end{equation}
where the first and second columns are respectively the vectors $\bf b_1$ and $\bf b_2$. The two matrices $\Hat T$ and $\Hat M$ are related by $\Hat M = \Hat T'^{-1}$ with $\Hat T'$ the transpose of $\Hat T$. 

For the later use, we here discuss the Fourier transform. The function $F(\vec R)$ defined on the honeycomb lattice sites can be expanded as 
\begin{equation}
F(\vec R) = \int_{\rm BZ}\frac{d\vec Q}{S_{\rm BZ}}F(\vec Q)e^{i\vec Q\cdot\vec R}, \label{F}
\end{equation}
where the $\vec Q$-integral is over the first BZ with $S_{\rm BZ}=2(2\pi)^2/\sqrt{3}$ its area, the components of $\vec R$ and $\vec Q$ are given in the orthogonal coordinate system, and $F(\vec Q)$ is the Fourier component of the function $F(\vec R)$. (A function and its Fourier component are distinguished by their arguments in this paper. We also adopt the convention that a capital vector implies its components defined in the orthogonal coordinate system, while a lowercase vector means that its components are given in the basis \{$\bf a_1,a_2$\} in real space or \{$\bf b_1,b_2$\} in the momentum space). In the basis \{$\bf a_1, a_2$\}, the function $F(\vec R)$ is given by $F(\Hat T\vec r) \equiv f(\vec r)$. From Eq. (\ref{F}), we get the expansion for the function $f(\vec r)$,
\begin{equation}
f(\vec r) = \int_{\rm BZ}\frac{d\vec q}{(2\pi)^2}F(\Hat M\vec q)e^{i\vec q\cdot\vec r}. 
\end{equation}
Therefore, the Fourier component of $f$ is given by $f(\vec q)=F(\Hat M\vec q)$. This relationship is useful for the Fourier transform of the Coulomb interaction.

\section{Tight-Binding Model}

For describing the electron system, we use the tight-binding model in which the nearest-neighbor hopping and the Coulomb interaction are taken into account. Firstly, we consider the hopping term,
\begin{equation} 
H_0 = -t\sum_{\langle ij \rangle\alpha}c^{\dagger}_{i\alpha}c_{j\alpha} \label{H0},
\end{equation}
where $t$ is the hopping parameter, $\langle ij \rangle$ means the nearest-neighbors, $c^{\dagger}_{i\alpha}$ ($c_{i\alpha}$) is the creation (annihilation) operator of electrons at site $i$ with spin $\alpha$. With the basis of \{$\bf a_1, a_2$\}, we hereafter designate the coordinates of an unit cell as that of the black site at the left lower corner of the cell as shown in Fig. \ref{fig1}. The position of a site can then be denoted as $(j,\mu)$ where $j$ implies $j$-th unit cell and $\mu = 1 (2)$ corresponds to black (green) site. The electron operator, for example the annihilation one, should be then denoted as $c_{j\alpha,\mu}$. Expanding the operator with the plane waves, we have
\begin{equation}
c_{j\alpha,\mu} = \frac{1}{N}\sum_{\vec k}c_{\vec k\alpha,\mu}\exp(i\vec k\cdot\vec r_j),
\end{equation}
with $N$ as the number of total unit cells of the lattice, and $\vec r_j$ as the position vector of the $j$-th unit cell. Under such a convention, we can easily rewrite $H_0$ in momentum space. The result is 
\begin{equation}
H_0 = \sum_{\vec k\alpha}\psi^{\dagger}_{\vec k\alpha}\Hat h_{\vec k}\psi_{\vec k\alpha},
\end{equation}
here the electron operators are given by spinors, $\psi^{\dagger}_{\vec k\alpha} = (c^{\dagger}_{\vec k\alpha,1},c^{\dagger}_{\vec k\alpha,2})$ with the first and second components denoting electrons respectively at the black and green sublattices, $\Hat h_{\vec k} = \epsilon_{\vec k 1}\sigma_1+\epsilon_{\vec k 2}\sigma_2$ with $\sigma$'s as the Pauli matrices, and
\begin{eqnarray}
\epsilon_{\vec k 1} &=& -t(1+\cos k_x + \cos k_y) \\ \label{ek1}
\epsilon_{\vec k 2} &=& -t(\sin k_x + \sin k_y) \label{ek2}
\end{eqnarray}
with $k_x$ and $k_y$ as the components of the electron momentum $\vec k$ in the basis of \{$\bf b_1, b_2$\}.

\begin{figure} 
\centerline{\epsfig{file=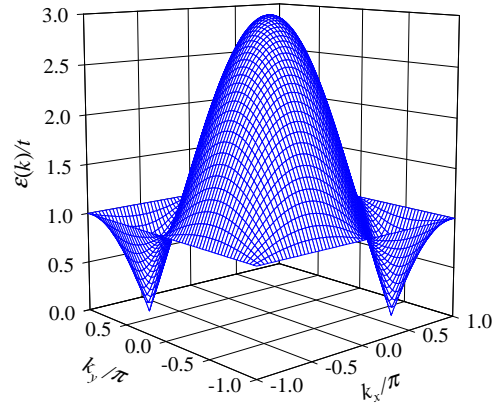, width=7. cm}}
\caption{(Color online) The energy of an electron in the upper band of non-interacting electrons as function of {$\bf k$}}. 
\label{fig3}
\end{figure}

The special feature of graphene is in the energy dispersion. The eigenvalues of $\Hat h_{\vec k}$ are $\pm \epsilon(\vec k)$ with 
\begin{equation}
\epsilon(\vec k) = \sqrt{\epsilon^2_{\vec k 1}+\epsilon^2_{\vec k 2}}.
\end{equation}
In Fig. \ref{fig3}, we show the energy dispersion of upper band. Clearly, $\epsilon(\vec k)$ depends on $\vec k$ linearly only when $\vec k$ closes to $\pm(1,-1)2\pi/3$ where it vanishes. In the continuous model, the energy dispersion is approximated by simple cones. 

We next consider the Coulomb interaction. The interaction $V_{\mu\nu} (\vec R_i,\vec R_j)$ between two electrons respectively at positions $(i,\mu)$ and $(j,\nu)$ is given by
\begin{eqnarray}
V_{\mu\mu} (\vec R_i,\vec R_j) &=& \begin{cases}\frac{e^2}{\epsilon|\vec R_i-\vec R_j|},~{\rm for}~ \vec R_i \ne \vec R_j\\
 U, ~{\rm for}~ \vec R_i = \vec R_j \end{cases}\\
V_{12} (\vec R_i,\vec R_j) &=& \frac{e^2}{\epsilon|\vec R_i-\vec R_j-\vec L|} \\
V_{21} (\vec R_i,\vec R_j) &=& \frac{e^2}{\epsilon|\vec R_i-\vec R_j+\vec L|} 
\end{eqnarray}
where $\vec R_{i(j)}$ denotes the position of the $i$ ($j$)-th unit cell, $\epsilon$ is the static dielectric constant due to the screening by the electrons of carbon core and the substrate, and $\vec L$ is the vector from the black site to green site in an unit cell as shown in Fig. 1. The on-site interaction $U$ is the Coulomb repulsion between electrons of anti-parallel spins, leading to the short-range antiferromagnetic correlations (AFC). Since the AFC is not significant in graphene, $U$ should not be too large. In our calculation, we set $U = 2e^2/\epsilon L$ which is double of the nearest-neighbor interaction. For the long-range Coulomb interacting system, the final result should not sensitively depend on such a small but reasonable $U$. By taking into account only the charge fluctuations (with the spin fluctuations neglected), the interaction term of the Hamiltonian in momentum space is given by
\begin{equation}
H_1 = \frac{1}{2N}\sum_{\vec q} n^{\dagger}_{\vec q}\Hat v_{\vec q} n_{\vec q}
\end{equation}
where $n^{\dagger}_{\vec q}=(n^{\dagger}_{\vec q1},n^{\dagger}_{\vec q2})$ is the electron density operator, and $\Hat v_{\vec q}$ is the Fourier component of the Coulomb interaction. Since the total charge of the electrons is neutralized by the background of positive charges, the $\vec q = 0$ term is excluded from the summation. The elements of $\Hat v_{\vec q}$ are given in Appendix A.

\section{Green's function}

\begin{figure} 
\centerline{\epsfig{file=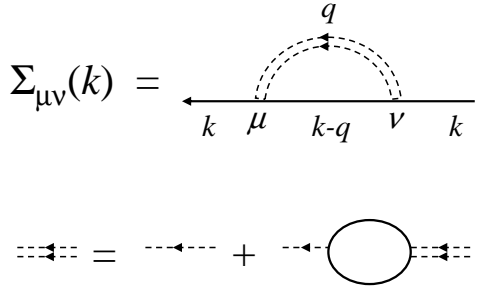,width=7.5 cm}}
\caption{(Color online) The electron self-energy in the renormalized-ring-diagram approximation. The solid lines are Green's functions, the single and double dashed lines are respectively the bare and renormalized Coulomb interactions, and the bubble is the electron polarizability.}\label{fig4}
\end{figure}

The Green function of electrons is defined as 
\begin{equation}
\Hat G(\vec k,\tau-\tau') = -\langle T_{\tau}\psi_{\vec k\alpha}(\tau)\psi^{\dagger}_{\vec k\alpha}(\tau')\rangle, 
\end{equation}
where $\tau$ is the imaginary time, and $\langle T_{\tau}\cdots\rangle$ means the statistical average of the $\tau$-ordered product of the operators. In the frequency space, $\Hat G$ is given by
\begin{equation}
\Hat G(\vec k,i\omega_n)= [i\omega_n-\Hat\xi_{\vec k}-\Hat\Sigma(\vec 
k,i\omega_n)]^{-1}, \label{G}
\end{equation}
where $\Hat\xi_k = \Hat h_k-\mu$ with $\mu$ the chemical potential, $\omega_n$ is the fermionic Matsubara frequency, and $\Hat\Sigma(\vec k,i\omega_n)$ is the self-energy. For brevity, we hereafter will use $k \equiv (\vec k, i\omega_n)$ for the arguments unless stated otherwise. Under the RRDA which is diagrammatically shown in Fig. \ref{fig4}, the elements of $\Hat\Sigma(k)$ are given by
\begin{equation}
\Sigma_{\mu\nu}(k)=-\frac{1}{N\beta}\sum_{q}G_{\mu\nu}(k-q)W_{\mu\nu}(q), \label{se}
\end{equation}
where $\beta = 1/T$ with $T$ as the temperature (in unit of $k_B=1$) of the system, $W_{\mu\nu}(q)$ is the element of the matrix of the screened interaction
\begin{equation}
\Hat W(q)= [1-\Hat V(q)\Hat\chi(q)]^{-1}\Hat V(q), 
\end{equation}
and the element of the polarizability $\Hat\chi(q)$ is given by
\begin{equation}
\chi_{\mu\nu}(q) = \frac{2}{N\beta}\sum_{k}G_{\mu\nu}(k+q)G_{\nu\mu}(k)  \label{chi} 
\end{equation}
with $q \equiv (\vec q, i\Omega_m)$ and $\Omega_m$ is the bosonic Matsubara frequency. The chemical potential $\mu$ is determined by the electron density (that is the electron number per site),
\begin{equation}
n = \frac{2}{N\beta}\sum_{k}G_{11}(k)\exp(i\omega_n0^+). \label{mu} 
\end{equation}
From equations (\ref{G})-(\ref{mu}), the Green functions can be self-consistently determined. Using the Pad\'e approximation,\cite{Vidberg} we obtain the retarded self-energy $\Hat \Sigma^r(\vec k,\omega)$ and then the Green function $\Hat G^r(\vec k,\omega)$ under the analytical continuation $i\omega_n \to \omega+i0^+$. 

In order to do an effective numerical calculation, it is necessary to get a clear understanding of the symmetry of the Green functions, and this is shown in Appendix B. 

On the other hand, we need to pay attention to the behavior of the screened interaction $\Hat W(q)$. Since it approaches to the bear Coulomb interaction $\Hat v(q)$ in the limit $i\Omega_m \to \infty$, $\Hat W(q)$ can be separated into two parts, $\Hat W(q) = \Hat v(q) + \Hat W_R(q)$, where $\Hat W_R(q)$ is the induced interaction from the ring diagrams. The contribution of $\Hat v(q)$ in the self-energy yields the exchange part. Here, an important point is that the behavior of $\Hat W_R(q)$ in the limit $\vec q \to 0$ by RRDA is very different from that by RPA. For the sake of illustration, we use the descriptions for $\Hat v(q)$ and $\Hat\chi(q)$ by their Pauli components. In the limit $\vec q \to 0$, we have 
\begin{eqnarray}
\Hat v(q) &\to& \frac{\Gamma}{Q}(1+\sigma_1)\nonumber\\
\Hat\chi(q) &\to& \chi_0(0,i\Omega_m)+\chi_1(0,i\Omega_m)\sigma_1 \nonumber
\end{eqnarray}
where $\Gamma = 4\pi e^2/\sqrt{3}\epsilon a$. For $\Hat W_R(q)$ in the same limit, we get
\begin{equation}
\Hat W_R(q) \to -\frac{q_m\Gamma}{Q(Q+q_m)}(1+\sigma_1)\label{walv}
\end{equation}
with $q_m = -2\Gamma[\chi_0(0,i\Omega_m)+\chi_1(0,i\Omega_m)]$. In RPA, $q_m$ vanishes for $m \ne 0$. However, in RRDA, $q_m \ne 0$, it means the dynamic screening effect in the long-wavelength interactions. 

As is well known, the ring diagrams in the RPA include the contribution of plasmon excitations. The plasmon frequency $\Omega_Q \propto \sqrt{Q}$ is determined by the long-wavelength behavior of the  charge polarizabilty $\tilde \chi_{RPA}(\vec q, i\Omega_m) \equiv [\chi_0(\vec q,i\Omega_m)+\chi_1(\vec q,i\Omega_m)]_{RPA} \propto Q^2$ at $\vec q \to 0$. This polarizability is calculated in the absence of interactions. Under the RRDA, the corresponding polarization diagram of $\chi_0(\vec q,i\Omega_m)+\chi_1(\vec q,i\Omega_m)$ is calculated with renormalized Green functions, and it does not have such a behavior as in RPA. The renormalized-ring-diagram summation does not result in the desired plasmon excitations. The correct way to obtain the plasmon excitations is to calculate the two-particle Green function in which the vortex corrections need to be considered. Under the RRDA that is a conserving approximation for the single-particle Green function, the kernel of the equation for the two-particle propagator is generated from the functional derivative of the self-energy diagrams with respect to the Green function.\cite{Baym} The calculation of the two-particle propagator needs a more complicated mathematical procedure and is beyond the scope of the present approach.

\section{Numerical method}

Under the present approximation, the screening effect is negligible only at sufficient large Matsubara frequencies. This requires a considerable amount of numerical calculations. To save the computer time, the summations over the Matsubara frequencies in Eqs. (\ref{se}) and (\ref{chi}) need to be performed with special method. We have developed a super-high-efficiency algorithm for the series summations.\cite{Yan} In the present calculation, we have used the parameters $[h,L,M] = [2,15,5]$ for selection of the Matsubara frequencies distributed in $L$ successively connected blocks each of them containing $M$ frequencies with $h$ the integer-parameter that the stride in the $\ell$-th block is $h^{(\ell-1)}$. The total number of the frequencies selected here is $L(M-1)+1 = 61$. The largest number $N_c$ is about $2^L(M-1) = 2^{17}$ for the cutoff frequencies $\Omega_{N_c}=2N_c\pi T$ and $\omega_{N_c} = (2N_c-1)\pi T$. For the lowest temperature considered here, $T/t = 0.01$, we have $\omega_{N_c}/t \sim 8235$. At the frequencies larger than the cutoff, $\Hat W_R(q)$ is negligible small. For illustrating the numerical method, an example for calculation of the element $\chi_{\mu\mu}(q)$ is given in Appendix C.

On the other hand, the momentum integrals in Eqs. (\ref{se}) and (\ref{chi}) are convolutions. These integrals can be efficiently carried out by Fourier transforms. Here again, we should pay attention to $\Hat W_R(q)$. It should be carefully transformed from $q$-space to $r$-space since it has a sharp peak at $\vec q = 0$ as indicated by Eq. (\ref{walv}). This long-wavelength singular part should be subtracted from $\Hat W_R(q)$ and can be treated especially. It saves computer time to perform the transform of this singular part in the orthogonal coordinate system because where it is isotropic and the $Q$-integral within a circle close to the origin can be reduced to a one-dimensional one. Only within this circle, we need a very fine mesh for the integral. In the rest part of the hexagon Brillouin zone, one can use a crude mesh doing the two-dimensional integral because where the integrand is not singular. The remained part of $\Hat W_R(q)$ should be a regular function except there may be some undulations close to $\vec q = 0$ due to the subtraction. 

\begin{figure} 
\centerline{\epsfig{file=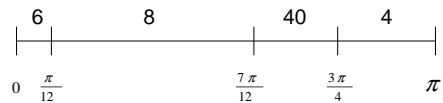,width=6.5 cm}}
\caption{For zero-doping calculation, the range $[0, \pi]$ is divided into 4 blocks with partition points $\pi/12, 7\pi/12$ and $3\pi/4$. Each number in the third line is the number of meshes in the corresponding block.}\label{mesh}
\end{figure}

Another point we should take care about is that the Green function varies drastically around and close to the Fermi surface at low temperatures. Since the band structure is not flat at the Fermi energy, the mesh for $q$-space integral should be fine enough around and close to the Fermi surface. We here give an example for sampling the points in momentum space. This example is for the zero-doping. Under the same consideration, the sampling for finite doping cases can be planed similarly. We divide the range $[0, \pi]$ in each axis in momentum space into four blocks shown as in Fig. \ref{mesh}. There are $N_i$ equal meshes in the $i$-th block and $N_i$'s are given by 
\begin{equation}
\{N_1,N_2,N_3,N_4\} = \{6,8,40,4\}. \nonumber
\end{equation}
The finest mesh is for the third block $[\frac{7\pi}{12},\frac{3\pi}{4}]$ which is centered with $2\pi/3$. In the Brillouin zone, this leads to a very fine mesh around the Dirac points $\pm(1,-1)2\pi/3$. The second finest mesh is for the first block, which is designed for dealing with the undulations of the remained part of $\Hat W_R(q)$ (after the subtraction of the sharp peak) when it is Fourier transformed from $q$-space to $r$-space. The rest two blocks have relatively crude meshes because the Green function is smooth there. 

Our numerical algorithm considerably saves the computer memory and time. The similar method and its accuracy have been demonstrated by a recent study on the two-dimensional electron system with infinite band width and long-range Coulomb interaction.\cite{Yan1} With our numerical algorithm, we have solved the above equations by iteration. 

\section{Results}

The system is characterized by the coupling constant $g$ that is defined by the ratio between the overall-average interaction energy $e^2/\epsilon a$ and the hopping energy $t$, 
\begin{equation}
g = \frac{e^2}{\epsilon at}.
\end{equation}
The parameters $t$ = 2.82 eV and $a$ = 2.4 {\AA} are known from the experimental observations.\cite{Bostwick} By choosing $\epsilon \sim 4$, we have $g \sim$ 0.5. Therefore, the graphene is a moderately-coupled Coulomb system.

\begin{figure} 
\centerline{\epsfig{file=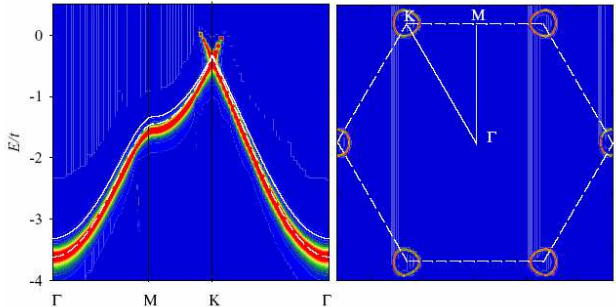,width=8.5 cm}}
\caption{(Color online) Left panel: Spectral density as a function of energy along the high symmetry directions in the Brillouin zone. Right panel: Spectral density at zero energy showing the Fermi surface structure. The doping concentration of the electrons is $c$ = 0.02, and the temperature is $T/t = 0.02$.}\label{fms}
\end{figure}

Firstly, we present the result for the spectral density that is defined by
\begin{eqnarray}
A(\vec k,E) &=& -\frac{1}{\pi}{\rm Im}{\rm Tr}\Hat G^r(\vec k,E) \nonumber\\
&=& -\frac{2}{\pi}{\rm Im}G^r_{11}(\vec k,E), \label{spect}
\end{eqnarray}
where $G^r_{11}(\vec k,E)=G^r_{22}(\vec k,E)$. In the non-interacting case, $A(\vec k,E)$ reduces to the $\delta$-functions representing the energy dispersions of the two bands. In the present case, the energy levels are broadened because of the many-body effect. Shown on the left panel in Fig. \ref{fms} is an intensity map of the spectral density in the energy-momentum plane at the doping concentration $c = n -1$ = 0.02 and temperature $T/t$ = 0.02. This map exhibits the energy distribution of the states as a function of momentum along the high symmetry directions $\Gamma$-M-K-$\Gamma$ in the Brillouin zone. For comparison, the free-particle energy dispersion $-\epsilon(\vec k)-\mu^0$ (with $\mu^0$ the chemical potential of the non-interacting system) is also depicted as the solid curve. Clearly, because of the Coulomb effect, the energy distribution has finite width, implying the finite life times of quasiparticle. In addition, the scale of the energy band is enlarged. The dashed curve is a rescale of the solid one. With comparing to the non-interacting dispersion, the energy band is magnified with a factor about 1.1. Actually, the exchange self-energy results in an additional hopping of the electrons, which renormalizes the kinetic energy.\cite{Yan2} To see this, we express the exchange self-energy in real space,
\begin{equation}
\Sigma^x_{ij} = -\langle c^{\dagger}_{j\alpha}c_{i\alpha}\rangle v_{ij}
\end{equation}
where $v_{ij}$ is the Coulomb interaction between electrons at sites $i$ and $j$. [Here $i$ and $j$ mean the positions of the lattice sites.] With the quantity $\Sigma^x_{ij}$, one can define the additional hopping parameter, $t^x_{ij} \equiv -\Sigma^x_{ij}$. The most sizable additional hopping is the one between the nearest-neighbor (NN). At $c$ = 0.02, and $g$ = 0.5, we find $t^x_{NN} = 0.22t$, and the magnitude of the next NN one is two order less than $t^x_{NN}$. At large distance, $t^x_{ij}$ is very small. As a total effect, such the additional hopping terms enlarge the original energy band. On the right panel of Fig. \ref{fms}, we show the map of the spectral density obtained by integration over the energy window of $0.06t$ around the chemical potential. The orbits of strong intensity correspond to the Fermi surfaces which are apparent not circles as compared with those from the simplified Dirac cone model. The structure of the Fermi surfaces is symmetrical under any rotation of angle $\pi/3$ around the origin. All these results are comparable with the ARPES experimental observations.\cite{Bostwick,Zhou} For $c$=0.0, we expect that the Fermi surfaces in Fig. \ref{fms} shrink into Dirac points.

The broadening of energy distribution of quasiparticle is described by the imaginary part of the self-energy. To be specific, we analyze the Green function $ G^r_{11}(\vec k,E)$. This function can be divided into two parts, $[G^r_u(\vec k,E)+G^r_l(\vec k,E)]/2$ with
\begin{eqnarray}
G^r_{u,l}(\vec k,E) = \frac{1}{E+\mu-\Sigma^r_0(\vec k,E)\mp S(\vec k,E)}\nonumber\\
S(\vec k,E) = \{[\epsilon_{\vec k,12}+\Sigma^r_{12}(\vec k,E)][\epsilon_{\vec k,21}+\Sigma^r_{21}(\vec k,E)\}^{1/2} \nonumber
\end{eqnarray}
where $\epsilon_{\vec k,12}=\epsilon^*_{\vec k,21} = \epsilon_{\vec k,1}-i\epsilon_{\vec k,2}$, $\Sigma^r_0(\vec k,E)\equiv \Sigma^r_{11}(\vec k,E)= \Sigma^r_{22}(\vec k,E)$, and $\Sigma^r_{\mu\nu}(\vec k,E)$'s are the components of the retarded self-energy matrix. The two Green functions $G_{u}$ and $G_l$ can be considered as for the electrons at the upper band and lower band, respectively. Corresponding to $G^r_{u,l}(\vec k,E)$, we define the retarded self-energies for the electrons respectively at upper and lower  bands as 
\begin{equation}
\Sigma^r_{u,l}(\vec k,E) = \Sigma^r_0(\vec k,E)\pm S(\vec k,E).
\end{equation}

\begin{figure} 
\centerline{\epsfig{file=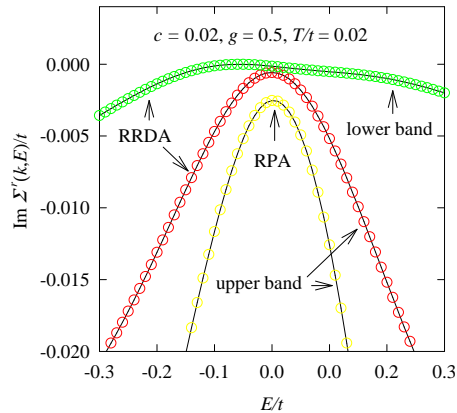,width=6.5 cm}}
\caption{(Color online) Imaginary part of the self-energy $\Sigma^r_{u,l}(\vec k,E)$ as function of $E$ at doping concentration $c$ = 0.02 and $T/t$ = 0.02 for the upper and lower bands. The result from the RPA for the upper band is also plotted here for comparison.}\label{se02}
\end{figure}

In Fig. \ref{se02}, the imaginary parts of the self-energies $\Sigma^r_{u,l}(\vec k,E)$ are presented as functions of $E$ at a Fermi momentum $\vec k \approx (-0.8\pi,0.6\pi)$ for the system at doping concentration $c$ = 0.02 and the temperature $T/t$ = 0.02. At small energy, ${\rm Im}\Sigma^r_{u}(\vec k,E)$ is a quadratic function of $E$. The value at $E$ = 0 is very small and should vanish at zero temperature. At a region of larger energy, ${\rm Im}\Sigma^r_{u}(\vec k,E)$ seems to be linearly depended on $E$. The magnitude of ${\rm Im}\Sigma^r_{l}(\vec k,E)$ is, however, small compared with ${\rm Im}\Sigma^r_{u}(\vec k,E)$. The reason is clear. At the Fermi surface, the states of energy $E$ is far from the corresponding states of the same momentum $\vec k$ at the lower band. Therefore, the self-energy $\Sigma^r_{l}(\vec k,E)$ is small. In Fig. \ref{se02}, the RPA result for the self-energy of upper band is also shown for comparison. The magnitude of ${\rm Im}\Sigma^r_u(\vec k,E)$ by RPA is larger than that by the RRDA. But, the quadratic $E$-dependence of ${\rm Im}\Sigma^r_u(\vec k,E)$ at small $E$ is also reflected by the RPA calculation. 

\begin{figure} 
\centerline{\epsfig{file=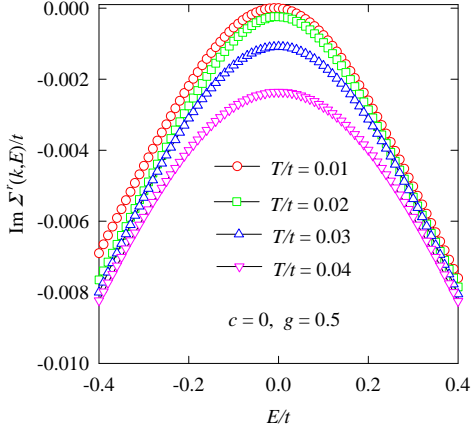,width=6.5 cm}}
\caption{(Color online) Imaginary part of the self-energy $\Sigma^r_{u}(\vec k,E)$ as function of $E$ at zero doping concentration for several different temperatures.}\label{se0}
\end{figure}

\begin{figure} 
\centerline{\epsfig{file=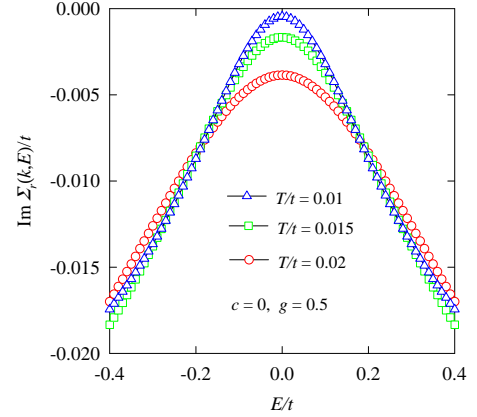,width=6.5 cm}}
\caption{(Color online) RPA result for the imaginary part of the self-energy $\Sigma^r_{u,l}(\vec k,E)$ as function of $E$ at zero doping concentration. }\label{rpase0}
\end{figure}

In Fig. \ref{se0}, the results for ${\rm Im}\Sigma^r_u(\vec k,E)$ are depicted at low temperatures and at zero doping concentration. Since each Fermi surface is a Dirac point in this case, the Fermi momentum for Fig. \ref{se0} is chosen as $\vec k = (-2\pi/3,2\pi/3)$. Firstly, at the Dirac point, because the state is the common state of the upper and lower bands, the self-energies of both bands coincide. At small energy, apart from a small value at $E$ = 0 due to the finite temperature effect, all the results for ${\rm Im}\Sigma^r_u(\vec k,E)$ show a quadratic dependence on $E$. Out of the small energy region, ${\rm Im}\Sigma^r_u(\vec k,E)$ linearly depends on $E$ to certain limit. For the purpose of comparison, The RPA results obtained from our approach at the zero doping concentration are exhibited in Fig. \ref{rpase0}. In the limit $T \to 0$, ${\rm Im}\Sigma^r_{u,l}(\vec k,E)$ appears to be linearly depended on $|E|$. This is in agreement with the analytical result obtained from the continuous model based on the RPA.\cite{Sarma} Therefore, the system at zero doping is referred \cite{Sarma} to  behave like a marginal Fermi liquid\cite{Varma} with the imaginary part of the self-energy going linear in energy near the chemical potential.

However, our numerical results based on the RRDA demonstrate that the imaginary parts of the self-energies fixed at the Fermi momentum are always varying as quadratic in energy close to the Fermi level, regardless the system is doped or not doped. This feature indicates that the quasiparticle in graphene behave like a moderately correlated Fermi liquid. 

\begin{figure} 
\vskip 5mm
\centerline{\epsfig{file=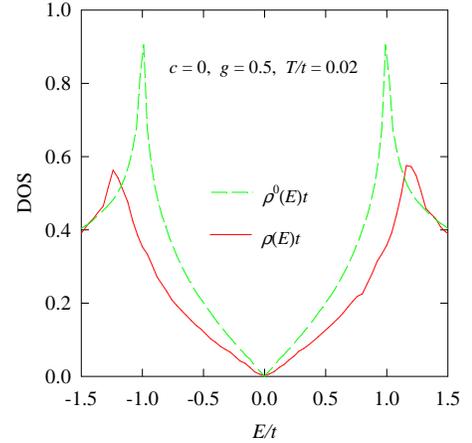,width=6.5 cm}}
\caption{(Color online) Density of states $\rho(E)$ as function of $E$ at zero doping concentration. $\rho^0(E)$ is the non-interacting counterpart.}\label{dos}
\end{figure}

With the retarded Green function, we can also calculate the density of states (DOS) defined as
\begin{equation}
\rho(E) = -\frac{2}{\pi N}\sum_{\vec k}{\rm Im}G^r_{11}(\vec k,E).
\end{equation}
Shown in Fig. \ref{dos} is the result for $\rho(E)$ at zero doping concentration and $T/t$ = 0.02. The non-interacting counterpart $\rho^0(E)$ is also depicted for comparison. Since the energy linearly depends on the magnitude of the momentum near the Dirac points, $\rho^0(E)$ is proportional to $E$ at small $E$. The two peaks come from the van Hove singularity because the energy bands are flat at $E = \pm t$ as shown in Fig. \ref{fig3}. Under the Coulomb interaction, the spectral density of quasiparticle is broadened. This results in lowering the density of states and smearing the peaks. With comparing to $\rho^0(E)$, the two peaks in $\rho(E)$ shift to larger energy because of the energy band enlarged by the exchange interaction. 
 
\begin{figure} 
\centerline{\epsfig{file=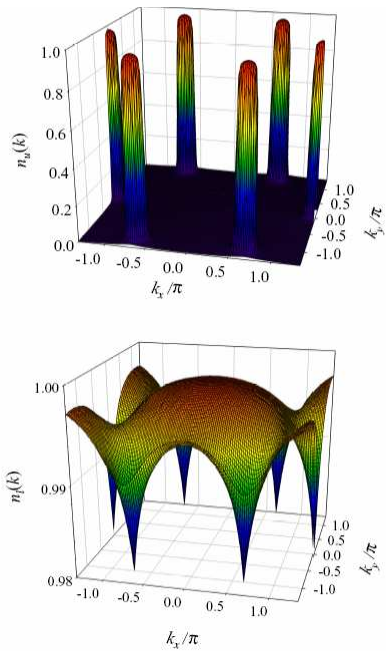,width=7. cm}}
\caption{(Color online) Distribution functions at doping concentration $c$ = 0.02 and $T/t = 0.02$. }\label{dist1}
\end{figure}

We next consider the distribution function $n(\vec k)$. For the graphene, $n(\vec k)$ should be defined as
\begin{equation}
n(\vec k) = \frac{1}{\beta}\sum_n{\rm Tr}\Hat G(\vec k,i\omega_n)\exp(i\omega_n0^+).
\end{equation}
As we have encountered in Eq. (\ref{spect}), $n(\vec k)$ is determined only by $G_{11}(\vec k,i\omega_n)$. Further more, since the Green function $G_{11}(\vec k,i\omega_n)$ can be divided into parts of the upper and lower bands, we therefore define the distribution functions for these two bands as
\begin{equation}
n_{u,l}(\vec k) = \frac{1}{\beta}\sum_nG_{u,l}(\vec k,i\omega_n)\exp(i\omega_n0^+).
\end{equation}
The total distribution function is given by $n(\vec k) = n_u(\vec k) + n_l(\vec k)$. In Fig. \ref{dist1}, we exhibit the result of the distribution functions $n_{u,l}(\vec k)$ at temperature $T/t$ = 0.02 and doping concentration $c$ = 0.02. In the inner Fermi area, the upper band is almost fully occupied with the electrons. The distribution $n_{u}(\vec k)$ drops drastically at the Fermi surface. The occupation at the lower band is not flat. Close to each Dirac point (the corner of the hexagon Brillouin zone), the behavior of $n_l(\vec k)$ looks like a sink. The depression of $n_l(\vec k)$ comes from two aspects. One is the temperature effect. The distribution $n_l(\vec k)$ reaches its minimum at the Dirac points because where the energy (Dirac energy) is the highest for states in the lower band. Since the doping concentration is small, the Dirac energy is close to the Fermi level. Therefore, $n_l(\vec k)$ has an apparent drop at the Dirac points. Another reason is due to the many-body effect. Because of the Coulomb interaction, an amount of electrons can be redistributed from the lower band to upper band.

\begin{figure} 
\centerline{\epsfig{file=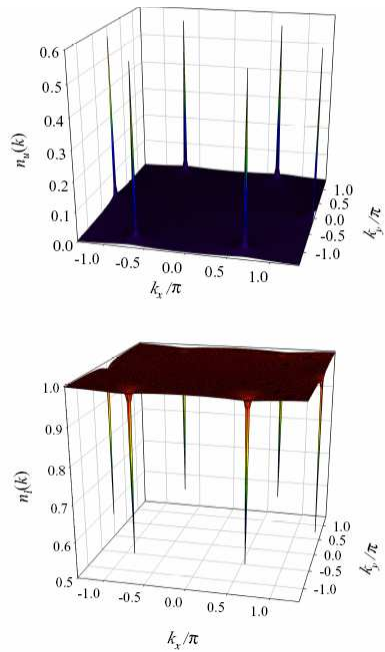, width=7. cm}}
\caption{(Color online) Distribution functions at zero doping concentration and $T/t = 0.02$. }\label{dist2}
\end{figure}

At zero doping, each Fermi surface shrinks to a point. The distribution functions are shown in Fig. \ref{dist2}. The distribution $n_u(\vec k)$ at upper band concentrates at the Dirac points. At lower band, there is a obvious depression of $n_l(\vec k)$ at the point. To see the many-body and temperature effects in the zero-doping case, we show the total distribution $n(\vec k)$ at and close to the Dirac points as functions of temperature. Very close to the Dirac points, $n(\vec k)$ increases drastically with decreasing temperature. At zero temperature, the total distribution at the Dirac points $1< n(\vec k_0) < 2$ can be expected. This is very different from the non-interacting distribution. The latter is constant 1 because the zero-energy levels of both bands are half occupied and at the momentum different from the Dirac points only the lower band is fully occupied. Under the Coulomb interactions, some of the electrons around each Dirac point in the lower band are gathered to the upper band close to the Dirac point, resulting in a higher distribution at and close to the point. From the increasing tendencies of $n(\vec k_0)$ and $n(\vec k_0+\Delta \vec k_0)$, we can infer there is an abrupt drop in $n(\vec k)$ close to each Dirac point. Therefore, the zero-doping distribution function at zero temperature is consistent with the Fermi liquid behavior.

For the negative doping, the Fermi surface opens again. In this case, the Fermi level is at the lower band. Since the upper and lower bands are symmetric about the zero energy, at zero temperature, the electron distribution at the upper band corresponds to the hole distribution at the lower band, and vice versa. Therefore, the distributions at negative doping can be obtained from that of positive doping. For example, at $c$ = -0.02, by flipping Fig. \ref{dist1} upside down, we obtain the image of distributions. The shape of each Fermi surface is the same as that at $c$ = 0.02. However, the Fermi area is outside of the surface where the lower band is near fully occupied.  

\begin{figure} 
\centerline{\epsfig{file=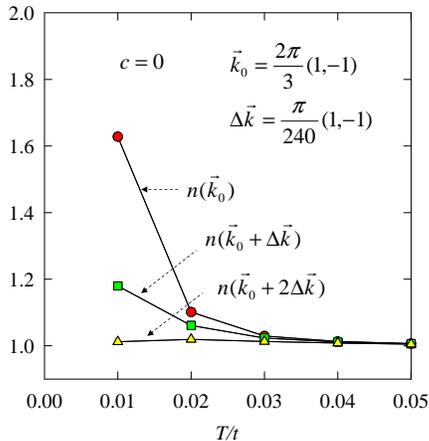, width=6.5 cm}}
\caption{(Color online) Temperature dependence of the total distribution at and close to the Dirac points at zero doping concentration. }\label{daf.ps}
\end{figure}

Finally, we give the ground-state energy per electron $\epsilon_0$. At low temperatures, $T/t < 0.1$, the numerical results for the energy per electron are almost a constant. By extrapolation, we then obtain $\epsilon_0$. The results for the two doping cases are, $\epsilon_0 = -1.10t$ for $c = 0$, and $\epsilon_0 = -1.09t$ for $c = 0.02$.

\section{Summary}

In summary, using the tight-binding model with long-range Coulomb interactions defined in a honeycomb lattice, we have presented the Green-function formulation for the electron in graphene. The interactions between electrons are treated with the renormalized-ring-diagram approximation. The integral equations for determining the Green function are solved self-consistently using our high-efficiency numerical algorithm. The obtained spectral densities are comparable with the experimental observations. Since the imaginary part of the self-energy of the electron Green function fixed at the Fermi momentum varies as quadratic in energy near the chemical potential for both doped and undoped systems, we conclude that electrons in graphene follow the Fermi liquid like behavior. In addition, we also calculated the density of states and the distribution functions. 

\acknowledgments

This work was supported by a grant from the Robert A. Welch Foundation under No. E-1146, the TCSUH, and the National Basic Research 973 Program of China under grant No. 2005CB623602.

\vskip 5mm
{\bf APPENDIX A: Fourier transform of Coulomb interaction between electrons on the honeycomb lattice}
\vskip 5mm

In this appendix, we present the Fourier transform of the Coulomb interaction between electrons on the honeycomb lattice. Because it is long-range interaction, the accurate result cannot be obtained from the direct summation by definition with numerical calculation. We must seek a fast converging scheme for summation over the lattice sites. To do this, we separate the interaction 
\begin{equation}
V(R) = \begin{cases}\frac{e^2}{\epsilon R},~{\rm for}~R\ne 0\\ 
  U,  ~{\rm for}~R=0
  \end{cases}
\end{equation}
into short-range and long-range parts, $V(R) = V_S(R) + V_L(R)$ with
\begin{equation}
V_L(R) = \frac{e^2}{\epsilon\sqrt{a^2+R^2}}, 
\end{equation}
where $a$ is a free parameter of positive quantity, and $V_S(R) = V(R)-V_L(R)$. At $R \gg a$, $V_S(R) = O(R^{-3})$, therefore it is a short-range interaction. Its Fourier transform can be obtained by the direct summation,
\begin{equation}
\tilde V_S(\vec Q) = \sum_j V_S(\vec R_j)e^{-i\vec Q\cdot\vec R_j}.  
\end{equation}
On the other hand, $V_L(R)$ is long ranged. Direct summation for the Fourier transform of it converges very slow. However, this function can be expressed as 
\begin{equation}
V_L(R) = \int\frac{d\vec Q}{(2\pi)^2}\phi(Q)e^{i\vec Q\cdot \vec R}  \label{VL}
\end{equation}
where the $\vec Q$-integral is over the whole momentum space, and 
\begin{equation}
\phi(Q)=\frac{2\pi e^2}{\epsilon Q}e^{-aQ}.
\end{equation}
The integral in Eq. (\ref{VL}) can be written in a summation over the integrals each of them over a Brillouin zone. Shifting all these Brillouin zones to the first Brillouin zone by the corresponding reciprocal lattice vectors $\vec Q_n$'s, we have
\begin{equation}
V_L(R) = \sum_n\int_{\rm BZ}\frac{d\vec Q}{(2\pi)^2}\phi(|\vec Q+\vec Q_n|)e^{i(\vec Q+\vec Q_n)\cdot \vec R} 
\end{equation}
The order of integral and summation can be changed. Express $\vec R$ as $\vec R = \vec R_j + \vec Z$ where $\vec R_j$ is the position of $j$-th unit cell of the honeycomb lattice, and $\vec Z$ is a vector within the unit cell. We get
\begin{equation}
V_L(|\vec R_j + \vec Z|) = \int_{\rm BZ}\frac{d\vec Q}{(2\pi)^2}\sum_n\phi(|\vec Q+\vec Q_n|)e^{i(\vec Q+\vec Q_n)\cdot \vec Z} e^{i\vec Q\cdot \vec R}
\end{equation}
where use of $\exp(i\vec Q_n\cdot\vec R_j) = 1$ has been made. From this equation, we recognize the Fourier transform of the function $F(\vec R_j,\vec Z)\equiv V_L(\vec R_j+ \vec Z)$ defined on the honeycomb lattice with $\vec Z$ a parameter, 
\begin{equation}
\tilde F(\vec Q,\vec Z) = \frac{2}{\sqrt{3}}\sum_n\phi(|\vec Q+\vec Q_n|)e^{i(\vec Q+\vec Q_n)\cdot\vec Z}
\end{equation}
where the factor  $2\sqrt{3}$ comes from the ratio between the area of the first Brillouin zone of the honeycomb lattice and that of the square lattice $(2\pi)^2$. Since $\phi(Q)$ decreases exponentially at large $|\vec Q_n|$, the $n$-summation converges very fast. 

We are now ready to express the elements of the interaction matrix $\Hat v(\vec q) \equiv \Hat v^S(\vec q) + \Hat v^L(\vec q)$. With the above functions, we have
\begin{eqnarray}
v^S_{11}(\vec q)&=& v^S_{22}(\vec q)= \tilde V_S(\Hat M\vec q),\\
v^S_{12}(\vec q)&=& v^{S*}_{21}(\vec q)= \sum_j V_S(\vec R_j-\vec L)e^{-i\Hat M\vec q\cdot\vec R_j},\\
v^L_{11}(\vec q)&=& v^L_{22}(\vec q)= \tilde F(\Hat M\vec q,0),\\
v^L_{12}(\vec q)&=& v^{L*}_{21}(\vec q)= \tilde F(\Hat M\vec q,-\vec L),
\end{eqnarray}
where $\Hat M\vec q$ maps the vector $\vec q$ under the basis $\{\bf b_1, b_2\}$ into the one in the orthogonal coordinate system. 

\vskip 5mm
{\bf APPENDIX B: Symmetry of the Green function defined on the honeycomb lattice}
\vskip 5mm

For doing numerical calculation, it is necessary to understand the symmetry of the Green function. We start the discussion with the definition of the Green function in real space,
\begin{equation}
G_{\mu\nu}(\vec r_i-\vec r_j,\tau-\tau')=-\langle T_{\tau}c_{i\alpha,\mu}(\tau)c^{\dagger}_{j\alpha,\nu}(\tau')\rangle, 
\end{equation}
which describes a particle of spin-$\alpha$ propagating from position $(j,\nu)$ to $(i,\mu)$. Firstly, this function has the following property,
\begin{equation}
G_{\mu\nu}(\vec r_i-\vec r_j,\tau-\tau')=G_{\nu\mu}(\vec r_j-\vec r_i,\tau-\tau')
\end{equation}
because that the configuration for a particle propagating from $(j,\nu)$ to $(i,\mu)$ is the same as in the inverse process. Therefore, the diagonal Green functions are even under $\vec r \to -\vec r$. Denoting $G_0 = G_{11} = G_{22}$, we obtain
\begin{equation}
G_0(\vec r,\tau)=G_0(-\vec r,\tau).
\end{equation}
On the other hand, for the off-diagonal part, we get
\begin{equation}
G_{12}(\vec r,\tau)=G_{21}(-\vec r,\tau).
\end{equation}
Though the off-diagonal Green functions have no definite parity, they can be separated into even and odd functions. For $G_{12}(\vec r,\tau)$, we have
\begin{equation}
G_{12}(\vec r,\tau)= G_1(\vec r,\tau)-iG_2(\vec r,\tau)
\end{equation}
where $G_1(\vec r,\tau)$ and $G_2(\vec r,\tau)$ are even and odd functions of $\vec r$, respectively. Now, we can express the Green function matrix in terms of Pauli matrix,
\begin{equation}
\Hat G(\vec r,\tau) = G_0(\vec r,\tau)\sigma_0 + G_1(\vec r,\tau)\sigma_1 + G_2(\vec r,\tau)\sigma_2
\end{equation}
where the Pauli components as functions of $\vec r$ have their definite parities. Obviously, in momentum space, as functions of $\vec k$, they have the same parities. Using the property of parity, the Fourier expansions of these components are given by
\begin{eqnarray}
G_{0,1}(\vec r,\tau) &=& \int_H\frac{d\vec k}{2\pi^2} G_{0,1}(\vec k,\tau) \cos(\vec k\cdot\vec r) \\
G_2(\vec r,\tau) &=& i\int_H\frac{d\vec k}{2\pi^2} G_2(\vec k,\tau) \sin(\vec k\cdot\vec r) 
\end{eqnarray}
where the $\vec k$-integrals are over the half Brillouin zone: $0 \le k_x \le \pi$ and $-\pi \le k_y \le \pi$. Further more, in the half Brillouin zone, we can separate the Green functions $G_{\sigma}(\vec k,\tau)$'s into even and odd functions of $k_y$, $G^{\pm}_{\sigma}(\vec k,\tau)$ with the superscripts $\pm$ denoting the parities,
\begin{equation}
G^{\pm}_{\sigma}(k_x,k_y,\tau)= [G_{\sigma}(k_x,k_y,\tau)\pm G_{\sigma}(k_x,-k_y,\tau)]/2.
\end{equation}
Correspondingly, in real space, $G_{\sigma}(\vec r,\tau)$'s are separated into even and odd functions of $r_y$. We have
\begin{eqnarray}
G^+_{0,1}(\vec r,\tau) &=& \int_I\frac{d\vec k}{\pi^2} G^+_{0,1}(\vec k,\tau) \cos(k_xr_x)\cos(k_yr_y) \nonumber \\
G^-_{0,1}(\vec r,\tau) &=& -\int_I\frac{d\vec k}{\pi^2} G^-_{0,1}(\vec k,\tau) \sin(k_xr_x)\sin(k_yr_y)  \nonumber\\ 
G^+_2(\vec r,\tau) &=& i\int_I\frac{d\vec k}{\pi^2} G^+_2(\vec k,\tau) \sin(k_xr_x)\cos(k_yr_y) \nonumber\\
G^-_2(\vec r,\tau) &=& i\int_I\frac{d\vec k}{\pi^2} G^-_2(\vec k,\tau) \cos (k_xr_x)\sin(k_yr_y) \nonumber
\end{eqnarray}
where the $\vec k$-integrals are now over the first quadrant of the first Brillouin zone: $0 \le k_x \le \pi$ and $0 \le k_y \le \pi$. 

We next consider the property of the Green functions under the exchange of the coordinates $(r_x,r_y) \to (r_y,r_x)$. Since the system is symmetric under this exchange, we have
\begin{equation}
G_{\mu\nu}(r_x,r_y,\tau)=G_{\mu\nu}(r_y,r_x,\tau).
\end{equation}
From the above definitions, one can obtain that the Pauli components $G^{\pm}_{0,1}$ are symmetric about the exchange but
\begin{eqnarray}
G^+_2(r_x,r_y,\tau) &=& G^-_2(r_y,r_x,\tau) \nonumber\\
G^-_2(r_x,r_y,\tau) &=& G^+_2(r_y,r_x,\tau). \nonumber
\end{eqnarray}

Making use of these symmetries in a numerical process, we need to calculate these Green functions only in half of the first quadrant in both real and momentum spaces. 

Finally, the symmetry related to the time reversion is 
\begin{equation}
G^{\pm *}_{\sigma}(\vec k,i\omega_n) = G^{\pm}_{\sigma}(\vec k,-i\omega_n) ~~~{\rm  for~ \sigma = 0,1,2}
\end{equation}
which is obtained from the definition.

The above discussion of symmetries for the Green functions applies to the Coulomb interactions, polarizabilities, and the self-energies as well. 

\vskip 5mm
{\bf APPENDIX C: An example for calculation of $\chi_{\mu\nu}(q)$}
\vskip 5mm

For the readers' convenience, we here give an example for calculation of $\chi_{\mu\nu}(q)$. For the briefness, we present only the result for $\chi_{\mu\nu}(q) \equiv \chi_0(q)$. The results for other elements and the method for calculation of the self-energy elements $\Sigma_{\mu\nu}(k)$ can be obtained immediately with the same consideration.
As has been mentioned in the main text, the momentum convolution of two Green functions in $\chi_0(q)$ can be evaluated by Fourier transform. In real space, it is given by
\begin{equation}
\chi_0(\vec r,i\Omega_m)=\frac{2}{\beta}\sum_nG_0(\vec r,i\omega_n+i\Omega_m)G_0(\vec r,i\omega_n)
\end{equation}
where the $n$-summation for the Matsubara frequencies is over $(-\infty,\infty)$. Using the property of the Green function, $G^*_0(\vec r,i\omega_n) = G_0(\vec r,-i\omega_n)$, we take the summation only for $n > 0$,
\begin{eqnarray}
\chi_0(\vec r,i\Omega_m)=\frac{4}{\beta}{\rm Re}\{\sum^{\infty}_{n=1}G_0(\vec r,i\omega_n+i\Omega_m)G_0(\vec r,i\omega_n) \nonumber\\
~~+ \sum^{[m/2]}_{n=1}G_0(\vec r,i\Omega_m-i\omega_n)G_0(\vec r,-i\omega_n) \nonumber\\
~~+ |G_0(\vec r,i\omega_{\bar n})|^2/2|^{\bar n = (m+1)/2}_{\rm if~m~is~odd}\}.\nonumber
\end{eqnarray}
where $[m/2]$ is the integer part of $m/2$. Applying our rule for the series summation, we get  
\begin{eqnarray}
\chi_0(\vec r,i\Omega_m)=\frac{4}{\beta}{\rm Re}\{\sum_{p}w_pG_0(\vec r,i\omega_{p}+i\Omega_m)G_0(\vec r,i\omega_{p}) \nonumber\\
~~+ \sum_{p}w^{[m/2]}_pG_0(\vec r,i\Omega_m-i\omega_{p})G_0(\vec r,-i\omega_{p}) \nonumber\\
~~+ |G_0(\vec r,i\omega_{\bar n})|^2/2|^{\bar n = (m+1)/2}_{\rm if~m~is~odd}\}\nonumber\\ 
~~+ \delta\chi_0(\vec r,i\Omega_m)\nonumber
\end{eqnarray}
where the summation is over the selected points $\{p\}$ with $w_p$ and $w^{[m/2]}_p$ the corresponding weights,\cite{Yan} and the last term is the contribution from terms beyond the cutoff $N_c$ for the Matsubara frequency,
\begin{equation}
\delta\chi_0(\vec r,i\Omega_m)=\begin{cases}{-\frac{2\delta_{r0}}{m\pi^2T}\sum\limits^{N_c+m}_{n=N_c+1}\frac{1}{2n-1}  ~~~~{\rm if~m > 0}}\\
{-\frac{4\delta_{r0}}{\pi^2T}[\frac{\pi^2}{8}-\sum\limits^{N_c}_{n=1}\frac{1}{(2n-1)^2}]  ~~~~{\rm if~m = 0}}\end{cases}.
\end{equation}
The present expression for $\chi_0$ is essentially the same as that in Appendix B of Ref. \onlinecite{Yan} except an additional factor 2 stemming from the degree of spin freedom in the present case and a misprinting in Ref. \onlinecite{Yan}. The notation of taking real part of the result should be assigned in the expression of $\chi$ in Appendix B of Ref. \onlinecite{Yan}..

\end{document}